# The structure of a recent nova shell as observed by ALMA

Marcos P. Diaz[1]★, Zulema Abraham[1], Valério A. R. M. Ribeiro[2,3,4], Pedro P. B. Beaklini[1] and Larissa Takeda[1]

[1]*Departamento de Astronomia, Instituto de Astronomia, Geofísica e Ciências Atmosféricas, Universidade de São Paulo (IAG/USP), Rua do Matão 1226, 05508-900, São Paulo, Brazil*
[2]*CIDMA, Departamento de Física, Universidade de Aveiro, Campus Universitário de Santiago, 3810-193 Aveiro, Portugal*
[3]*Instituto de Telecomunicações, Campus Universitário de Santiago, 3810-193 Aveiro, Portugal*
[4]*Department of Astrophysics/IMAPP, Radboud University, P.O. Box 9010, 6500 GL Nijmegen, The Netherlands*

**ABSTRACT**
High resolution ALMA observations of the recent (2.52 yrs old) shell of Nova V5668 Sgr (2015) show a highly structured ionised gas distribution with small ($10^{15}$ cm) clumps. These are the smallest structures ever observed in the remnant of a stellar thermonuclear explosion. No extended contiguous emission could be found above the 2.5 $\sigma$ level in our data, while the peak hydrogen densities in the clumps reach $10^6$ cm$^{-3}$. The millimetre continuum image suggests that large scale structures previously distinguished in other recent nova shells may result from the distribution of bright unresolved condensations.

## 1 INTRODUCTION

Following hypernovae and supernovae, novae (classical and recurrent) are the third most energetic stellar explosive phenomenon. They are the result of the building up of a thin hydrogen rich surface layer on the surface of a white dwarf by mass transfer in a semi-detached binary system. This layer ignites a thermonuclear runaway to produce a transient source radiating near the Eddington luminosity (Starrfield et al. 1974). The life of the host compact star is closely regulated by the secular companion mass loss and irradiation due to the nova outburst cycles (Nelemans et al. 2016). The white dwarf mass growth time-scale is comparable to the companion nuclear time-scale. Therefore, the occurrence of the single degenerate channel to SN Ia from cataclysmic variables is closely related to the mass ejection in nova outbursts (see Maoz et al. (2014) for a recent review). Being found in the Galaxy at relatively high rates (Mróz et al. 2015; Shafter 2017), novae are key to the understanding of degenerate thermonuclear runaways and mass ejection phenomena in a variety of astrophysical scenarios. On the other hand, nova progenitors, as an old binary population, have long term yields from their explosive runaways, contributing to the abundances of Lithium (Tajitsu et al. 2016) and other isotopes of special astrophysical interest such as Na$^{22}$ (Sallaska et al. 2011) and Al$^{26}$ (Bennett et al. 2013).

Understanding the nature of nova shells and their evolutionary consequences relies on the knowledge of the mass distribution in the ejecta. Photoionisation models of shells assuming homogeneous symmetric components have lead to biased results (Moraes & Diaz 2011) while two-dimensional and multi-wavelength observations are essential to constrain the physical and chemical diagnostic. The presence of significant density contrasts was first proposed from the wide ionisation range and optically thick [OI] lines seen in nova optical spectra (Williams 1994). Imaging of old shells showing condensations (Shara et al. 1997; Schaefer et al. 2010), polar caps and belts (Moraes & Diaz 2009) added to the complex scenario of shells and their current 3D modelling using adaptive optics feed 2D spectroscopy (Takeda et al. 2018). While the clumpy nature of shells is essential to explain the observed spectrum in the optical and X-rays (Williams 2013), the difficulties of resolving and following structures in young nova remnants have contributed to the poor understanding of their formation and evolution within the ejecta. The unprecedented angular resolution and sensitivity at sub-mm wavelengths afforded by ALMA, opens a new paradigm in the study of stellar outburst debris.

Nova Sgr 2015b (V5668 Sgr) was one of the brightest nova of this century, reaching a peak visual magnitude of 4.3. This rare D-type nova was also one of the best observed, both from ground and space, in an extended multi-wavelength effort (see Gehrz et al. (2018) for a review).

★ E-mail: marcos.diaz@iag.usp.br

2     *M. P. Diaz et al.*

## 2 OBSERVATIONS

The ALMA observations of V5668 Sgr were made in band 6 (215.4 - 232.8 GHz) during cycle 4, in 2017, September 28.0 UT and lasted 70 min with a 34-min exposure on target. A total of 42 antennas were employed with baselines ranging from 60 m to 12.7 km, the largest current baseline configuration (C40-9), which allows the contiguous mapping of 1 arcsec FOV. Three 2 GHz spectral windows, centred at 229.2, 219.0 and 216.5 GHz, were used for continuum observation while a fourth window was centred at 231.9 GHz, the frequency of the H30$\alpha$ recombination line. The latter has a 1.875 GHz band width divided in 3840 channels, providing a resolution of 0.63 km s$^{-1}$.

The Hogboom algorithm implemented in the CASA software (McMullin et al. 2007) was used to obtain the continuum image in two different ways. First, we used it with Briggs weighting, to recover the best possible resolution; then, we applied the natural weighting, which gives lower resolution, to evaluate and validate the properties of the Briggs weighted reconstruction. The images were constructed on a 256×256 pixels grid with 5 mas pixel size and cleaned until they reach 0.026 mJy beam$^{-1}$ (RMS). Due to the weakness of the source, we were not able to self-calibrate in either weightings. For the Briggs weighted image, the maximum flux density in the map is 0.16 mJy beam$^{-1}$ with a final RMS noise of 0.021 mJy beam$^{-1}$, recovering a beam size of 31 × 26 mas at PA = 75°. For the natural weighting the resulting beam is of 60 × 40 mas at PA = 85°, with similar peak value and RMS noise. Both weighting methods provide comparable (<12% difference) integrated flux densities.

We followed the same strategy to obtain the image of the line emission, by cleaning with different velocity resolutions up to 6.3 km s$^{-1}$. The H30$\alpha$ recombination line was not detected in the data cube, using either Briggs or natural weighting. We were able to clean each velocity channel to 0.1 mJy beam$^{-1}$ (RMS), which may be taken as an upper limit to the line flux density. The recombination line flux estimated in LTE using the emission measure from sec. 3 falls below our detection limit.

## 3 RESULTS

The continuum emission of the shell at 230 GHz is shown in fig. 1a. This is the best resolution image calculated from our data. Figure 1b displays the best resolution image contours superposed to the natural weighting, lower resolution map, which confirms the structures seen in the best resolution image. An even lower resolution image, aimed to match the Hubble Space Telescope (HST) Point Spread Function (PSF) at H$\alpha$, simulates the expected emission of an optically thin recombination line (fig. 1c). This image, derived directly from the high resolution image (fig. 1a) using a gaussian kernel convolution, is remarkably similar to the ring/bipolar shell morphology proposed by Harvey et al. (2018) for this nova. Most if not all the large scale structure seen in fig. 1c is due to the blending of a distribution of smaller unresolved sources. Large contiguous emission regions that could be identified as a ring or bipolar caps, are absent above the 2.5 $\sigma$ level in ALMA high resolution image, although they may be present at fainter flux densities. The total measured shell flux density is 9.4 ± 1.5 mJy.

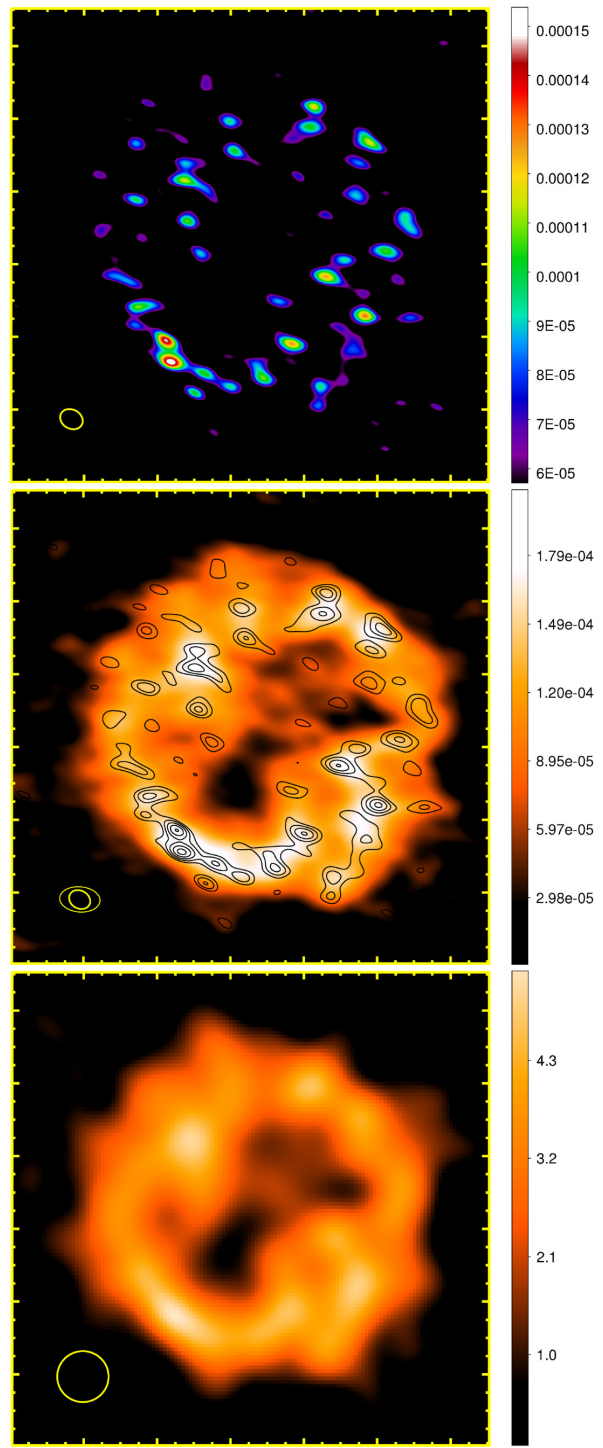

**Figure 1.** ALMA image of the V5668 Sgr shell at 230 GHz. The top panel (a) shows the full resolution continuum image above 3.0 $\sigma$. Major tick marks are spaced by 100 mas. The middle panel (b) shows the natural weighted image with 5 superposed contour levels from the image in panel (a), starting at 2.5 $\sigma$. The bottom panel (c) displays the image given in panel (a) after a gaussian convolution to match the HST spatial resolution at H$\alpha$. The flux unit is Jy beam$^{-1}$ for panels (a) and (b), and arbitrary units for panel (c). North is at the top and East to the left.

The power spectrum of the continuum image (fig. 2) shows that a significant fraction of the emission is formed in spatially resolved structures. It also indicates that features above the noise level are present at scales as small as 55 mas or $1\times10^{15}$ cm, i.e. near the resolution limit of the observations. A distance of 1540 pc is assumed in the present work (Banerjee et al. 2016), a value that is in agreement with both the observed shell expansion parallax for $V_{exp} = 650$ km $s^{-1}$ and MMRD (maximum magnitude and rate of decay) relations. No parallax to this object could be found in GAIA DR2. Measurement of individual clump sizes in the image are consistent with the small scales found in the power spectrum analysis. As expected, a noise dominated power spectrum is obtained beyond the beam size up to the Nyquist frequency.

Nova V5668 Sgr formed dust soon after eruption (Banerjee et al. 2016). In order to determine the total dust thermal emission the detailed grain data for this nova (Gehrz et al. 2018) was used, extrapolating the grain temperature and assuming a constant central source luminosity. An upper limit to the dust emission of <0.2 mJy at 230 GHz was found, which corresponds to <3% of the total observed flux. A continued dust destruction after day 112 was also found (Gehrz et al. 2018), which decreases the upper limit quoted above. Therefore, the main continuum emission process is identified as thermal bremsstrahlung. Regions above 2.5 $\sigma$ level were contoured and 32 emission maximums within those regions were identified and measured. The flux density at these maximums correlate (R = 0.6) with the size of its corresponding condensation (fig. 3), suggesting an optically thin free-free emission. Astrometric flux-weighted centroids and free-free peak densities were derived for the expanding clumps (material available upon request). The condensation peak density distribution has a single maximum at $N_e \sim 7\times10^5$ cm$^{-3}$, assuming $N_e/N_H = 2$ and a temperature $T_e = 8000$ K. The estimated densities indicate an optically thin free-free regime at 230 GHz. A constraint to the overall shell filling factor $f$ < 0.2 may be derived from the condensation measured sizes and upper limits, together with the observed outer shell radius of 230 mas. This upper limit to $f$ is smaller than the values often considered in the modelling of unresolved novae. The hydrogen mass depleted in the ionised clumps is estimated as $\sim 7\times 10^{-6}$ M$_\odot$.

## 4 DISCUSSION AND CONCLUSIONS

The behaviour of a nova at millimetre wavelengths is similar to those at centimetre wavelengths (Ivison et al. 1993; Bode & Evans 2008), where in both cases the continuum imaging observations commonly trace the free-free thermal emission. A contribution from synchrotron radiation at centimetre range may be found a few months after outburst (e.g. V959 Mon, Chomiuk et al. (2014)), which is expected to be much weaker at sub-millimetre wavelengths. The classical view is that as the ejecta surface area expands the flux density increases during the optically thick phase. Once the ejecta becomes optically thin, first at the lowest wavelengths, the flux density decreases as the photosphere recedes. Our simple optical depth calculations suggest that the millime-tre observations presented here are already in the optically thin phase. The perception of small structures in nova shells using HST imaging and ground based adaptive optics raised the problem of their unknown size distribution. The smallest structure sizes quoted here could be compared to model turbulence and Rayleigh-Taylor instability scales (Toraskar et al. 2013) and also to the optically thick nova fireball. While the role of density instabilities and radiative shocks (Metzger et al. 2014; Derdzinski et al. 2017) in the formation of cooler density enhancements is still a matter of debate, small condensations are not expected to survive against photoevaporation (Bertoldi & McKee 1990; Mellema et al. 1998) when partially ionised, given the observed density gradients and the irradiation initially provided by the nova photosphere followed by a reestablished accretion disc.

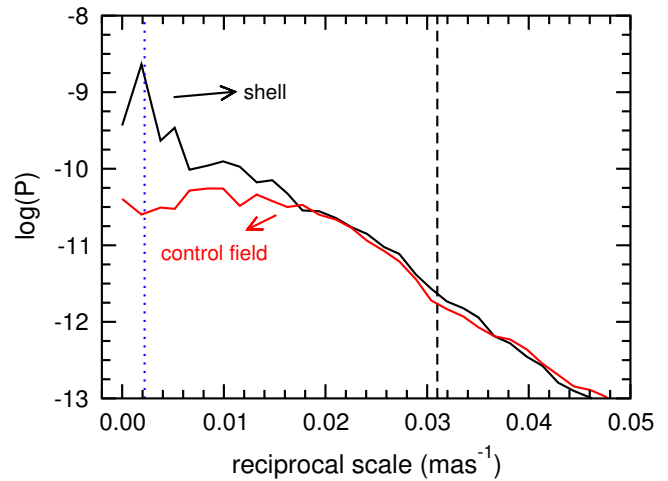

**Figure 2.** Power spectrum of ALMA image from fig. 1a. The black curve shows the unfiltered power spectrum of a tapered circular section containing the nova shell. The red curve was derived from a pure noise region subjected to the same procedure. The vertical lines indicates the beam maximum size (right) and the whole shell scale (left).

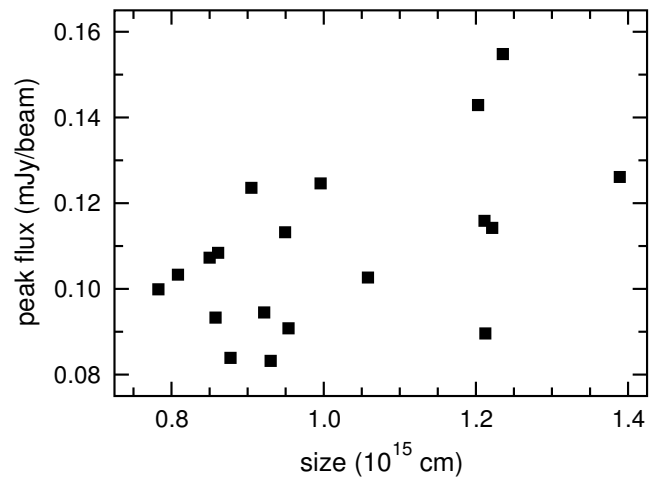

**Figure 3.** Peak continuum flux density vs. average projected size at 2.5 $\sigma$ level for a sub-sample of resolved and marginally resolved clumps.

4    *M. P. Diaz et al.*

The selection of a nearby recent nova provided the early detection of very small structures. The remnant power spectrum suggests that the structure scales may extend towards even smaller knots, unresolved by current ALMA capabilities. The unresolved imaging of compact condensations may be the case of other recent nova shells observed at lower spatial resolution.

**ACKNOWLEDGEMENTS**

This paper makes use of the following ALMA data: ADS/JAO.ALMA# 2016.1.00682.S. ALMA is a partnership of ESO (representing its member states), NSF (USA) and NINS (Japan), together with NRC (Canada), MOST and ASIAA (Taiwan), and KASI (Republic of Korea), in cooperation with the Republic of Chile. The Joint ALMA Observatory is operated by ESO, AUI/NRAO and NAOJ. MPD thanks CNPq grant 305657. ZA grants: CNPq 305768/2015-8 and FAPESP 2015/50360. We also acknowledge FAPESP fellowship and visitor support 2014/10326-3, 2014/07460-0 and 2015/16489-4. VARMR acknowledges financial support from the Radboud Excellence Initiative and Fundação para a Ciência e a Tecnologia (FCT) in the form of an exploratory project of reference IF/00498/2015, from Center for Research & Development in Mathematics and Applications (CIDMA) strategic project UID/MAT/04106/2013 and supported by Enabling Green E-science for the Square Kilometer Array Research Infrastructure (ENGAGESKA), POCI-01-0145-FEDER-022217, funded by Programa Operacional Competitividade e Internacionalização (COMPETE 2020) and FCT, Portugal. This work has made use of computing facilities of the Laboratory of Astroinformatics (IAG/USP, NAT/Unicsul), whose purchase was made possible by the Brazilian agency FAPESP (grant 2009/54006-4) and the INCT-A. Finally, we would like to thank the anonymous referee for the improving comments on the manuscript.

This paper has been typeset from a TeX/LaTeX file prepared by the author.